# ON THE INTERFERENCE FROM VDE-SAT DOWNLINK TO THE INCUMBENT LAND MOBILE SYSTEM

Michael Wang

## 1. Introduction

Although both frequency utilization plans for VDE-SAT downlink are within the frequency range of 156.0125 – 162.0375 MHz which belongs to the VHF maritime mobile band (from 156 to 174 MHz, inclusive), <u>on land</u>, this band is mostly allocated for conventional and trunked land mobile systems by safety agencies, utilities and transportation companies, e.g., police, fire, ambulance services, dispatched services. Many industries use land mobile service as their primary means of communication, especially from a fixed location to a fleet of mobile stations. Therefore, the potential interference from the VDE-SAT systems to the incumbent land mobile systems in these frequency bands must be taken into consideration for protection against the harmful interference to these land mobile services due to the global broadcast nature of a satellite system.

The challenge lies in the fact that there is no existing regulatory rule directly established for protection of the land system against the satellite system or the like, and hence the evaluation of the potential impact on the incumbent land systems becomes difficult if not impossible. A method adopted in the current analysis is to place general restrictions on the emissions from the satellite stations inferred from the existing regulatory rules for interference protection <u>between</u> land systems specified by ITU and ECC. The restrictions are expressed in terms of values of maximum allowed *power flux density* (PFD) emitted by any space stations to the surface of the Earth at all possible incident angles in a reference bandwidth that serves as a "protection mask" for the land system such that the *actual* interference that the land system experiences is no worse than that from another land mobile system permitted by these regulations.

We first take a look at two simplest regulatory constraints specified by ITU and ECC in these bands, and its implications on the PFD mask.

## 2. Interference-to-Noise Constraint

ITU defines a simple protection criteria of interference-to-noise ratio for the conventional and trunked land mobile systems operating below 890 MHz. It requires that the interference power seen by the receiver of the land mobile systems be at least 6 dB below the noise level of the receiver, i.e.,

$$\frac{I}{N} \leq \mu^{\text{ITU-I/N}}, \qquad (1)$$

where $\mu^{\text{ITU-I/N}} = -6$ dB. Here, $N$ is the noise power of the receiver system, which consists of both internal noise and external noise [1]. The internal noise is dominated by the thermal noise with the noise temperature of $T_s = 30$ dBK, given the receiver noise figure of 7 dB. For VHF maritime mobile band, galactic noise with temperature of $T_g = 24$ dBK and man-made noise with temperature in the range from $T_m = 31$ to $38$ dBK are the main source of external noise. The overall equipment temperature is thus,

$$T = T_s + T_g + T_m \text{ (linear)} \approx 34 \text{ dBK} \qquad (2)$$

assuming $T_m = 31$ dBK.

Given the receiver channel bandwidth of 15 kHz for the legacy digital land mobile stations, the system noise level is

$$N = \kappa TB = -153 \text{ dBW}, \tag{3}$$

where $\kappa$ is Boltzmann's constant (i.e., $1.38 \times 10^{-24}$ J/K). The maximum interference seen by the receiver is thus

$$I_{\max}^{\text{ITU-I/N}} = \mu^{\text{ITU-I/N}} \cdot N = -159 \text{ dBW} \tag{4}$$

per 15 kHz.

In particular, for a typical VDE-SAT space station antenna with circular polarization, the signal incident on the antenna (linear polarization) of the land undergoes a $\mathcal{G}_{\text{land}}^{\text{VDE-SAT}} = 3$ dB mismatch loss before entering the receiver. The allowed maximum interference impinging on the antenna of a land system from a space station is thus

$$I_{\max}^{\text{ITU-I/N}} = -159 \text{ dBW} + \mathcal{G}_{\text{land}}^{\text{VDE-SAT}} \text{dB} = -156 \text{ dBW} \tag{5}$$

per 15 kHz.

For systems operating below the 15 GHz band, the reference bandwidth of 4 kHz is appropriate when considering the impact of unwanted signals at the input of terrestrial station receivers of an ITU hypothetical reference circuit. The interference in (5) corresponds to

$$\begin{aligned} I_{\max}^{\text{ITU-I/N}} &= -156 \text{ dB}(\text{W}/15\text{kHz}) + 6 \text{ dBkHz} \\ &= -162 \text{ dBW,} \end{aligned} \tag{6}$$

per 4 kHz (6 dBkHz).

It is clear that the actual interference that a land receiver sees depends on the angle of incidence, $\theta$, relative to the antenna direction, which is also the angle relative to the horizon since land mobile antennas are normally omnidirectional and directed toward the horizon. When the incoming interference is line with the antenna direction, i.e., $\theta = 0°$, the interference received by the land mobile reaches to maximum, which means that the land mobile is capable of tolerating greater interfering PFD at other angles giving the maximum interference, i.e.,

$$\Theta^{\text{ITU-I/N}}(\theta) = 4\pi\lambda^{-2}\varepsilon I_{\max}^{\text{ITU-I/N}} G_{\text{land}}^{-1}(\theta), \tag{7}$$

where $\lambda$ is the wavelength of the interference, $\varepsilon$ the receiver feeder loss, and $G_{\text{land}}(\theta)$ the gain of receiving antenna. For the land mobile station an average antenna gain pattern (relative to an isotropic antenna) is considered, and the resulting antenna patterns with respective peak antenna gain $G_{\text{land}}(0) = 8.15$ dBi and 2.15 dBi for mobile and base stations are depicted in the Fig. 1. The corresponding PFDs for mobiles and base stations are denoted $\Theta_{\text{mobile}}^{\text{ITU-I/N}}(\theta)$ and $\Theta_{\text{base}}^{\text{ITU-I/N}}(\theta)$, respectively. In this case, $\theta$ is also the land station to satellite elevation angle above the horizon.

Clearly, the ultimate interference PFD mask inferred from the ITU interference-to-noise ratio constraint is

$$\tilde{\Theta}^{\text{ITU-I/N}}(\theta) = \min\left\{\Theta_{\text{base}}^{\text{ITU-I/N}}(\theta), \Theta_{\text{mobile}}^{\text{ITU-I/N}}(\theta)\right\}, \theta \in [0°, 90°]. \tag{8}$$

## 3. Field Strength Constraint

The ECC recommendation T/R 25-08 also provides five indicative *coordination thresholds* with a reference bandwidth of 25 kHz for interference avoidance among land mobile systems from neighboring countries in the frequency band from 29.7 to 470 MHz. For the maritime mobile frequency bands, the corresponding coordination field strength $E$ is 12 dB(μV/m) per 25 kHz, or 4 dB(μV/m) per 4 kHz. It translates to a PFD value of

$$\Theta^{\text{ECC}} = \frac{E^2}{120\pi} = -142 \ (\text{dBW/m}^2) \tag{9}$$

per 4 kHz. If we take into consideration the polarization loss between the space station antenna and the land station antenna, the maximum allowed PFD is

$$\Theta^{\text{ECC}}(\theta) = \Theta^{\text{ECC}} G_{\text{land}}(0) G_{\text{land}}^{-1}(\theta) + \vartheta_{\text{land}}^{\text{VDE-SAT}} \, \text{dB} \tag{10}$$

per 4 kHz for both cases of mobiles $\Theta_{\text{mobile}}^{\text{ECC}}(\theta)$, and base station $\Theta_{\text{base}}^{\text{ECC}}(\theta)$. It is apparent that the case for mobiles is more stringent than the base stations, and hence we have the effective PFD mask of

$$\tilde{\Theta}^{\text{ECC}}(\theta) = \min\left\{\Theta_{\text{base}}^{\text{ECC}}(\theta), \Theta_{\text{mobile}}^{\text{ECC}}(\theta)\right\} = \Theta_{\text{mobile}}^{\text{ECC}}(\theta), \ \theta \in \left[0°, \ 90°\right]. \tag{11}$$

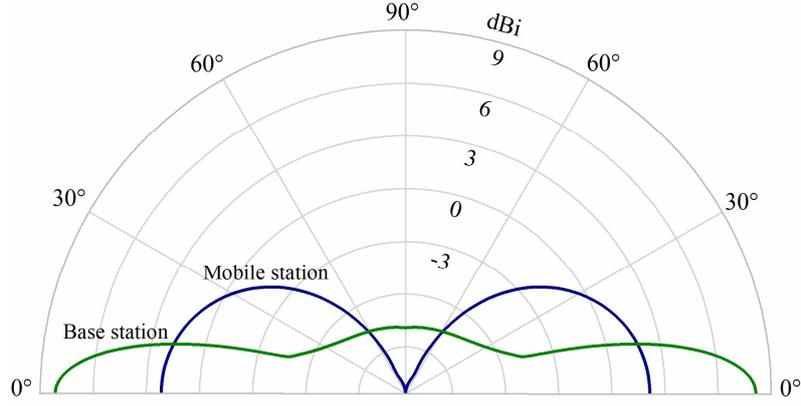

Fig. 1. Illustration of antenna gain $G_{\text{land}}(\theta)$ as a function of elevation angle $\theta$ for both land mobile and base station cases.

This is the maximum allowed PFD emitted by a VDE-SAT satellite on the surface of the Earth inferred from the ECC coordination threshold, and is plotted in Fig. 2 together with the PFD mask deduced from the ITU interference-to-noise ratio constraint.

It is observed that there exists a 20-dB gap between the PFD masks derived from the two criteria recommended by ITU and ECC. Although simple, the interference-to-noise ratio criterion produces a more stringent PFD mask that seems to be over-protective.

Fortunately, there exists another protection criterion developed by ITU from the perspective of the ultimate system performance requirements which is more relevant and adequate.

## 4. Carrier-to-Interference Constraint

For digital land mobile systems in the frequency band between 138 and 174 MHz, a bit error rate of 2-5% with C4FM modulation is targeted. The corresponding required energy per bit to noise plus interference power spectral density ratio, $E_b/N_0 \approx 10$ dB, is specified in [3], which translates to a minimum carrier-to-noise plus interference ratio of

Table 1 Typical parameters of land mobile systems.

| Land station type | Base station | | Mobile station | |
|---|---|---|---|---|
| | Digital | Analog | Digital | Analog |
| Modulation type | C4FM | FM | C4FM | FM |
| Output power (typical/min) | 60W/20W | 30W/5W | 30W/1W | |
| Antenna gain | 8 dBi | | 2 dBi | |
| Feeder loss | 2 dB | | 1 dB | |
| Antenna height | 65 m | | 2 m | |
| Distance to horizon | 29 km | | 5 km | |
| Free space loss | 107 dB | | | |
| Additional path loss | 34 dB | | | |
| Channel bandwidth | 15 kHz | | | |

$$\varsigma_\mathrm{d} \triangleq \frac{E_\mathrm{b}}{N_0} \cdot \frac{2R_\mathrm{s}}{B} \approx 13 \text{ dB}, \tag{12}$$

where $R_\mathrm{s}$ is modulation symbol rate and $B$ the channel bandwidth, i.e.,

$$\frac{C}{N+I} \geq \varsigma_\mathrm{d} \tag{13}$$

or

$$\frac{C}{I} \geq \left( \varsigma_\mathrm{d}^{-1} - \left( \frac{C}{N} \right)^{-1} \right)^{-1}, \tag{14}$$

where $C$ is the received carrier power, $N+I = N_0 B$ is the noise plus interference power, and $N$ is given in (3).

The performance of an analog system is commonly measured by the ratio of the total received power to the unwanted power (SINAD). A typical SINAD value of $\varsigma_\mathrm{a} = 12$ dB for establishing degradation protection for analogue land mobile systems is defined as

$$\frac{C+N+I+D}{N+I+D} \geq \varsigma_\mathrm{a} \tag{15}$$

or

$$\frac{C}{I} \geq \left( (\varsigma_\mathrm{a} - 1)^{-1} - \left( \left( \frac{C}{N} \right)^{-1} + \left( \frac{C}{D} \right)^{-1} \right) \right)^{-1}, \tag{16}$$

where $C/D$ is the signal to distortion ratio, typically in the vicinity of 20 dB [4].

According to the representative parameters of technical and operational characteristics of conventional and trunked land mobile systems operating in the frequency band 156.0125- 162.0375 MHz listed in Table 1, the typical antenna height of base stations and mobile stations are 65 m and 2 m, respectively, thus the maximum line-of-sight distance between the base station and mobile station is $d = 29 \text{ km} + 5 \text{ km} = 34 \text{ km}$, which leads to a free space loss of

$$\ell = \left( \frac{4\pi d}{\lambda} \right)^2 \approx 107 \text{ dB} \tag{17}$$

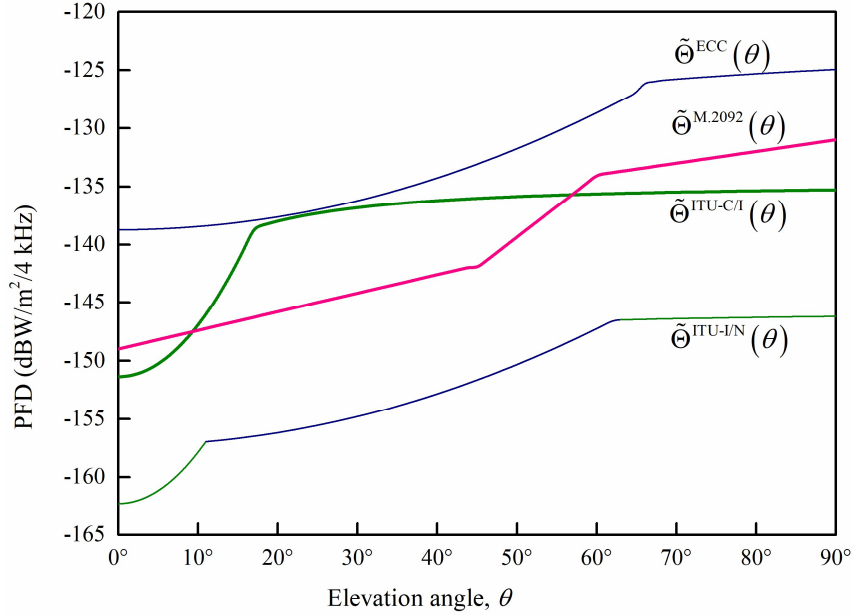

Fig. 2 PFD masks deduced from ECC and the two ITU constraints, together with the PFD mask, $\tilde{\Theta}^{M.2092}(\theta)$, specified in Recommendation ITU-R M.2092-0.

in the VHF maritime mobile frequency band. Noting that in addition to free space loss, land mobile station transmission channel also experiences an excess path loss which is estimated as [5]

$$\Delta \ell \approx 34 \text{ dB}, \tag{18}$$

giving rise to a total path loss of

$$L_{\text{land}} = \ell + \Delta \ell \approx 141 \text{ dB}. \tag{19}$$

After taking into account the receiver feeder loss, the minimum carrier power received by the land mobile station receiver, i.e., the sensitivity is

$$C_{\min} = \frac{\Lambda_{\text{land}}(0) \cdot G_{\text{land}}(0) \cdot \varepsilon^{-1}}{L_{\text{land}}}, \tag{20}$$

where $\Lambda_{\text{land}}(0)$ is the minimum effective isotropically radiated power (EIRP) of the land station transmitter at an elevation angle of $\theta = 0°$, recalling that land mobile service transceivers direct to each other at near $0°$ elevation angle (the horizon). Note that the EIRP is defined as the product of transmit power and transmit antenna gain (including the feeder loss).

Referring back to the system noise level $N$ in (3), for the digital land system, the corresponding minimum carrier to interference ratio is

$$\zeta^{\text{ITU-d}} \triangleq \left( \varsigma_d^{-1} - \left( \frac{C_{\min}}{N} \right)^{-1} \right)^{-1}, \tag{21}$$

and the maximum allowed interference level follows

$$I_{\max}^{\text{ITU-d}} = \frac{C_{\min}}{\zeta^{\text{ITU-d}}}, \tag{22}$$

which is $-134$ dBW and $-148$ dBW per 15 kHz for land mobile and base stations, respectively. Similarly, for the analog land system, we define

$$\zeta^{\text{ITU-a}} \triangleq \left( (\varsigma_a - 1)^{-1} - \left( \left( \frac{C_{\min}}{N} \right)^{-1} + \left( \frac{C_{\min}}{D} \right)^{-1} \right) \right)^{-1}, \quad (23)$$

and the maximum allowed interference is therefore

$$I_{\max}^{\text{ITU-a}} = \frac{C_{\min}}{\zeta^{\text{ITU-a}}}, \quad (24)$$

which is $-139$ dBW and $-147$ dBW per 15 kHz for land mobile and base stations, respectively.

Combining the digital and analog scenarios, and taking the polarization loss into consideration, we have

$$I_{\max}^{\text{ITU-C/I}} = \min\{ I_{\max}^{\text{ITU-d}}, I_{\max}^{\text{ITU-a}} \} + \vartheta_{\text{land}}^{\text{VDE-SAT}} \ (\text{dB}), \quad (25)$$

i.e., $-136$ dBW per 15 kHz for mobile stations, and $-145$ dBW per 15 kHz for base stations.

Similar to (7) the PFD constraint inferred from the performance based requirement is

$$\Theta^{\text{ITU-C/I}}(\theta) = 4\pi \lambda^{-2} \varepsilon I_{\max}^{\text{ITU-C/I}} G_{\text{land}}^{-1}(\theta). \quad (26)$$

The mask on PFD irradiating the surface of the Earth is thus

$$\tilde{\Theta}^{\text{ITU-C/I}}(\theta) = \min\{ \Theta_{\text{mobile}}^{\text{ITU-C/I}}(\theta), \Theta_{\text{base}}^{\text{ITU-C/I}}(\theta) \}, \quad \theta \in [0°, 90°], \quad (27)$$

which is plotted in Fig. 2 in $\text{dBW}/\text{m}^2$ per 4 kHz. It is seen that this PFD mask agrees better with the one from the ECC coordination threshold, and more importantly it reflects the thing (i.e., the performance) that matters most.

The PFD mask specified in the recommendation ITU-R M.2092-0 is also plotted in Fig. 2 for comparison.

## 5. Conclusion

The PFD mask derived from the ITU Carrier-to-Interference Constraint is a more adequate PFD mask for protection of land mobile systems against interference from satellite systems. The PFD mask specified in the recommendation ITU-R M.2092-0 agrees reasonably well with this mask except at low and high elevation angles.

`